\documentclass[aps,prl,showpacs,twocolumn,groupedaddress]{revtex4}

\usepackage{graphics}
\usepackage{epsfig}
\usepackage{mathrsfs}
\usepackage{amsbsy}
\usepackage{amssymb}
\usepackage{setspace}

\newcommand{\half}{\mbox{$\frac12$}}

\begin{document}

\title{On Constructing the Minimal Seed of Turbulence:\\ Nonlinear Transient Growth}

%\title{Towards Optimal Turbulence-triggering Disturbances in
%  Shear Flows via Nonlinear Transient Growth}
%\title{Transient Growth in Shear Flows:\\Linearity vs Nonlinearity}

% resubmission iteration 2nd July  RRK
% resubmission iteration 10th July  CCTP
% resubmission iteration 15th July RRK

\author{Chris C.T. Pringle}
\email{C.C.T.Pringle@reading.ac.uk}
\author{Rich R. Kerswell}
\email{R.R.Kerswell@bristol.ac.uk}

\affiliation{Department of Mathematics, University of Bristol, University Walk,
   Bristol BS8 1TW, United Kingdom}
\date{\today}

\begin{abstract}

Linear transient growth analysis is commonly used to suggest the
structure of disturbances which are particularly efficient in
triggering transition to turbulence in shear flows. We demonstrate that
the addition of nonlinearity to the analysis can substantially change
the prediction made in pipe flow from simple 2 dimensional streamwise
rolls to a spanwise and cross-stream localised 3 dimensional
state. This new nonlinear optimal is demonstrably more efficient in
triggering turbulence than the linear optimal indicating that there
are better ways to design perturbations to achieve transition.

\end{abstract}

\maketitle

Shear flows are ubiquitous in nature and engineering, and
understanding how and why they become turbulent has huge economic
implications.  This has led to a number of simplified canonical
problems being studied such as plane Couette flow, channel flow and
pipe flow which commonly exhibit turbulent behavior even when the
underlying laminar state is linearly stable.  In this case, a finite
amplitude perturbation is required in order to trigger turbulence and
a leading question is then what is the `most dangerous' or `smallest' 
such perturbation (with the metric typically being energy). 
Beyond its intrinsic interest, such information is fundamentally
important for devising effective control strategies to delay the
onset of turbulence.  

Linear transient growth analysis has commonly been used to suggest the
structure of such dangerous disturbances \cite{gustav}.  The basic
premise being that disturbances which experience the most (transient)
growth are also most efficient at modifying the underlying shear to
produce instability and probable transition to turbulence
\cite{zikanov,reddy}.  Recent improvements in our understanding of the
laminar-turbulent boundary or `edge', which determines whether a given
initial condition leads to a turbulent episode or to the laminar
state, has supported this idea albeit extended to disturbances of
finite amplitude. This is because some regions of this boundary have a much
smaller energy level (e.g. \cite{viswanath}) than the attracting
region of the boundary-confined dynamics \cite{itano} so that the minimum
energy point on the edge (the most dangerous disturbance) must
experience considerable energy growth as it sweeps up to the
attracting plateau.

However, there are two tacit assumptions in using linear transient
growth to identify critical disturbances: (1) that the energy growth
experienced by the most dangerous disturbance is the largest (or near
largest) possible at the critical energy; and (2) the optimal
disturbance which emerges from linear transient growth analysis
reasonably approximates the finite-amplitude optimal for the properly
nonlinear growth calculation. Efforts to test these assumptions have
concentrated on very small dimensional systems \cite{manneville}
or restricted the search for dangerous disturbances within small
subspaces \cite{viswanath,reddy} with broadly supportive results.  The
only study to specifically test assumption (2) used the Blasius
approximation for the boundary layer \cite{zuccher} and found no qualitative
difference between the nonlinear and linear optimals.

In this letter, however, we show for the first time using the full
Navier-Stokes equations how nonlinearity can fundamentally change the
optimal which emerges from a transient growth analysis in pipe flow at
subcritical energy levels, thereby contradicting assumption (2).  The
significance of this new state is that: a) it provides a much more
efficient way to trigger transition than the linear optimal; and b) it
is 3-dimensional and shows signs of localisation thereby appearing
more physically relevant than the 2-dimensional streamwise-independent
linear optimal.

The transient growth problem is the optimisation question: what
initial condition $\mathbf{u}(\mathbf{x},t=0)$ (added as a
perturbation to the laminar flow) for the
governing Navier-Stokes equations with fixed (perturbation) kinetic
energy $E_0$ will give rise to the largest subsequent energy $E_T$ at
a time $t=T$ later. This corresponds to maximising the functional
\begin{eqnarray}
\mathscr{L} &:=& \langle \half
\mathbf{u}(\mathbf{x},T)^2\rangle-\lambda\langle \half
\mathbf{u}(\mathbf{x},0)^2-E_0\rangle \nonumber \\ 
&&-\int_0^T \langle \boldsymbol{\nu},\bigg[\frac{\partial \mathbf{u}}{\partial t} 
-16su\hat{\mathbf{z}}+2(1-4s^2)\frac{\partial \mathbf{u}}{\partial z} \nonumber\\
&& \qquad \qquad \qquad +\mathbf{u}\cdot\nabla\mathbf{u}
+ \nabla p - \frac{1}{Re} \nabla^2 \mathbf{u} \bigg] \rangle dt
\nonumber \\
&&-\int_0^T \langle \Pi \nabla \cdot \mathbf{u} \rangle dt 
-\int_0^T \Gamma \langle \mathbf{u} \cdot \mathbf{\hat{z}} \rangle dt
\end{eqnarray}
where $\langle \, \, \rangle$ represents volume integration;
$(s,\phi,z)$ are cylindrical coordinates directed along the pipe;
$\lambda$, $\boldsymbol{\nu}(\mathbf{x},t)$, $\Pi(\mathbf{x},t)$ and
$\Gamma(t)$ are Lagrange multipliers imposing the constraints of
initial energy $E_0$, that the Navier-Stokes equations hold over $t
\in [0,T]$, incompressibility and constant mass flux in time
respectively (the system has been non-dimensionalised by the pipe
diameter $D$ and the bulk velocity $U$ so that $Re:=\rho UD/\mu$ where
$\rho$ is the density and $\mu$ is the dynamic viscosity, and the 
laminar flow is $2(1-4s^2)\hat{\mathbf{z}}\,$). Vanishing of the
variational derivatives requires that $\mathbf{u}$ must evolve
according to the Navier-Stokes equations, $\boldsymbol{\nu}$ evolves
according to the adjoint-Navier-Stokes equations and at times $t=0$
and $T$ we have optimality and compatibility conditions linking the
two sets of variables (e.g. see \cite{guegan} for details of the
linearised problem).  The method of solution is one of iteration as
follows:
\begin{itemize}
%\addtolength{\itemsep}{-0.75\baselineskip}
\item Make an initial guess for $\mathbf{u}(\mathbf{x},t=0)$ and allow the flow to evolve according to the
  Navier-Stokes equations until $t=T$.
\item Solve the compatibility condition for
  $\boldsymbol{\nu}(\mathbf{x},T)$, $ \delta \mathscr{L}/\delta
  \mathbf{u}(\mathbf{x},T) \equiv
  \mathbf{u}(\mathbf{x},T)-\boldsymbol{\nu}(\mathbf{x},T) =
  \mathbf{0}$.
\item Allow the incompressible field $\boldsymbol{\nu}(\mathbf{x},t)$
  to evolve \emph{backwards} in time until $t=0$ via the
  adjoint-Navier-Stokes equations
\begin{eqnarray}
&& \frac{\partial \boldsymbol{\nu}}{\partial t} + 2(1-4s^2)\frac{\partial \boldsymbol{\nu}}{\partial s} 
+ \frac{1}{s}(\nu_\phi u_s - \nu_s u_\phi)\boldsymbol{\hat{\phi}}  
 + \mathbf{u} \cdot \nabla) \boldsymbol{\nu} \nonumber \\ 
&&\qquad + 16s\nu_3\mathbf{\hat{s}}+(u_i \partial_j \nu_i)
= -\nabla \Pi - \frac{1}{Re} \nabla^2 \boldsymbol{\nu}
\end{eqnarray} 
\item Move $\mathbf{u}(\mathbf{x},0)$ in the direction of the
  variational derivative $ \delta \mathscr{L}/\delta
  \mathbf{u}(\mathbf{x},0) \equiv
  -\lambda\mathbf{u}(\mathbf{x},0)+\boldsymbol{\nu}(\mathbf{x},0)$ to
  increase $\mathscr{L}$ and repeat.
\end{itemize}
The algorithm should converge if $E_0$ does not exceed the critical
energy for transition. Beyond this, sensitivity to initial conditions
when $\mathbf{u}(\mathbf{x},T)$ reaches the turbulent state will lead
to non-smoothness. 

Both direct and adjoint equations were solved using
a fully spectral, primitive variables approach.  Time stepping was
done using a second order fractional step scheme, checked carefully
against the code of \cite{willis}. The computational domain was a
short periodic domain of length $\pi$ radii with typical spatial
resolution of 29 real Fourier modes azimuthally, 11 real Fourier modes
axially and 25 modified Chebyshev polynomials radially in each of the
8 scalar fields $(u,v,w,p,\nu_1,\nu_2,\nu_3,\Pi)$.  All
results have been checked for robustness to resolution changes.
Retention of the nonlinear terms poses a fresh technical challenge:
although the adjoint equation is linear in $\boldsymbol{\nu}$, it is
dependent on the evolution history of the forward variable
$\mathbf{u}$ which now must be stored.

\begin{figure}
% \begin{center}
\centering
  \includegraphics[width=\columnwidth]{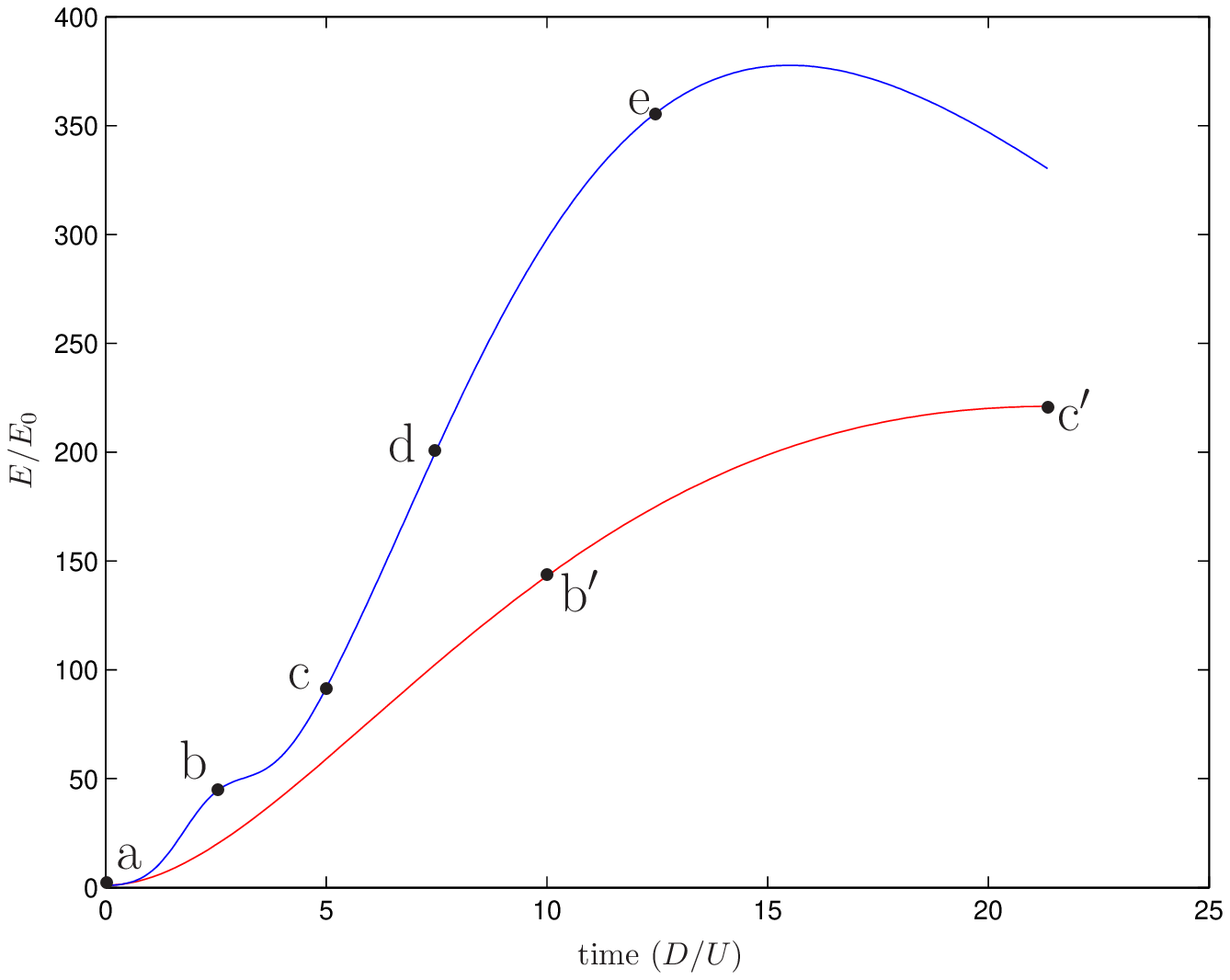}
  \caption{The evolution of the linear and nonlinear optimals at
    $Re=1750$. The blue (upper) line corresponds to the nonlinear optimal for
    $E_0=2\times 10^{-5}$ while the red (lower) line is the linear optimal
    ($E_0 \rightarrow 0$). The nonlinear result produces more growth
    and actually reaches its maximum at a slightly earlier time than
    $T$. \label{nl_opt}}
    
% \end{center}
\end{figure}

%
% Results
%

The linear transient growth optimal $\mathbf{u}_{lin}(\mathbf{x};Re)$
in pipe flow is well-known to be streamwise-independent (2D) rolls
which evolve into much larger streamwise-independent streaks
\cite{gustav}: see figs 1 and 2. Maximum growth occurs at
$T_{lin}\approx 12.2 \times Re/1000\,(D/U)$ \cite{mes}.  Introducing
nonlinearity (ie increasing $E_0$ from $0$), setting $T=12.2 \times
Re/1000\,(D/U)$ and allowing only 2D flows, leads smoothly to a
modified 2D optimal $\mathbf{u}_{2D}(\mathbf{x};E_0,Re)$ with
monotonically decreasing growth consistent with previous simulations
\cite{zikanov}.  Opening the optimisation up to fully 3D flows
initially just recovers the 2D result but once $E_0$ crosses a small
threshold $E_{3D}$ ($1.35 \times 10^{-5}$ at $Re=1750$), a completely
new optimal $\mathbf{u}_{3D}(\mathbf{x};E_0,Re)$ appears.  This 3D
optimal emerges from the optimisation procedure after it initially
appears to converge to the 2D optimal and then transiently visits an
intermediate state: see fig 3. Identifying this `loss of stability' of the 2D
optimal provided an efficient way to compute $E_{3D}(Re)$.  All
optimisation results were robust over three very different choices of
starting flow: a) $\mathbf{u}_{lin}$ with noise; b) the asymmetric
travelling wave \cite{PK07}; and c) a turbulent flow snapshot (all
rescaled to the appropriate initial energy). This supported our supposition
that the algorithm samples all possible flows of a given energy to
select the global optimiser although no proof is available.

\begin{figure}
% \begin{center}
\centering
  \includegraphics[width=\columnwidth]{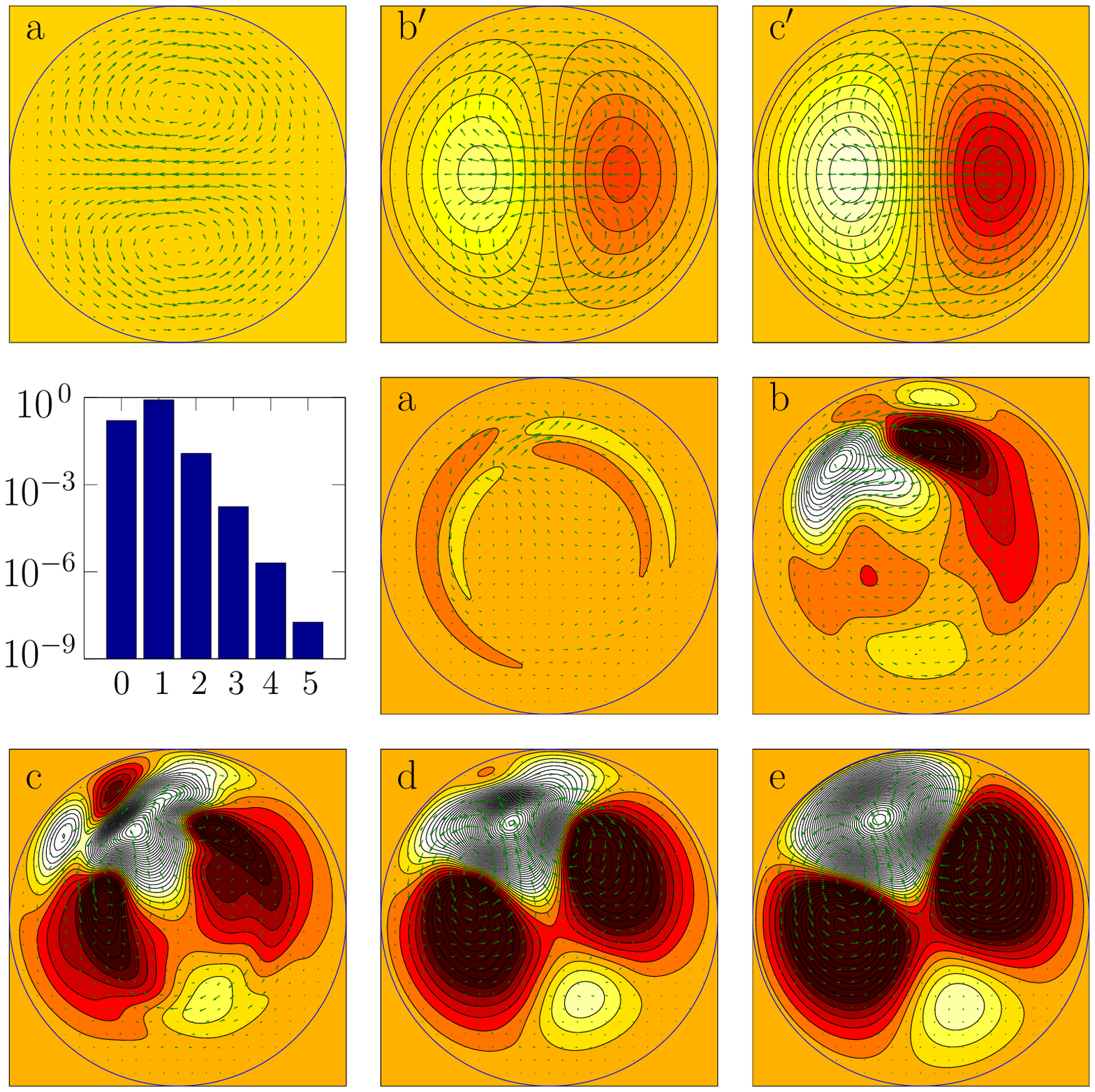}
  \caption{Three snapshots of the linear optimal (top) and five
    snapshots (middle \& bottom) of the 3D optimal for $Re=1750$ and
    $E_0=2\times 10^{-5}$ during its evolution.  Labels refer to
    figure \ref{nl_opt}, arrows indicate cross-sectional velocities
    and colours axial velocity beyond the laminar flow (white/light
    for positive and red/dark for negative: outside shade represents
    zero). The bar chart shows the ratio of energy in each 
    streamwise Fourier mode of the initial nonlinear optimal (a).
    \label{nl_evol}}
% \end{center}
\end{figure}

Given the intensity of the runs ($O(200)$ iterations and each
iteration requires integrating forwards and backwards over the period
$[0,T]$), one other Reynolds number, $Re=2250$, was selected to
confirm our findings. Here, the new 3D state becomes the nonlinear
optimal at $E_{3D}=4.8\times 10^{-6}$ and has essentially the same
appearance as at $Re=1750$: see fig. \ref{nl_evol}.  Unlike the linear
optimal which is globally simple in form and undergoes an evolution
that is well established (rolls advecting the mean shear to generate
streaks), the 3D optimal is localised to one side of the pipe and
initially has both rolls and streaks of comparable amplitude. Figures
\ref{nl_opt} and \ref{nl_evol} show a new 2-stage evolution: 
a preliminary phase when the flow delocalises
followed by a longer growth phase where the flow structure stabilises
to essentially two 2D large-scale slow streaks sandwiching one fast
streak near the boundary.

\begin{figure}

% \begin{center}
\centering

  \includegraphics[width=\columnwidth]{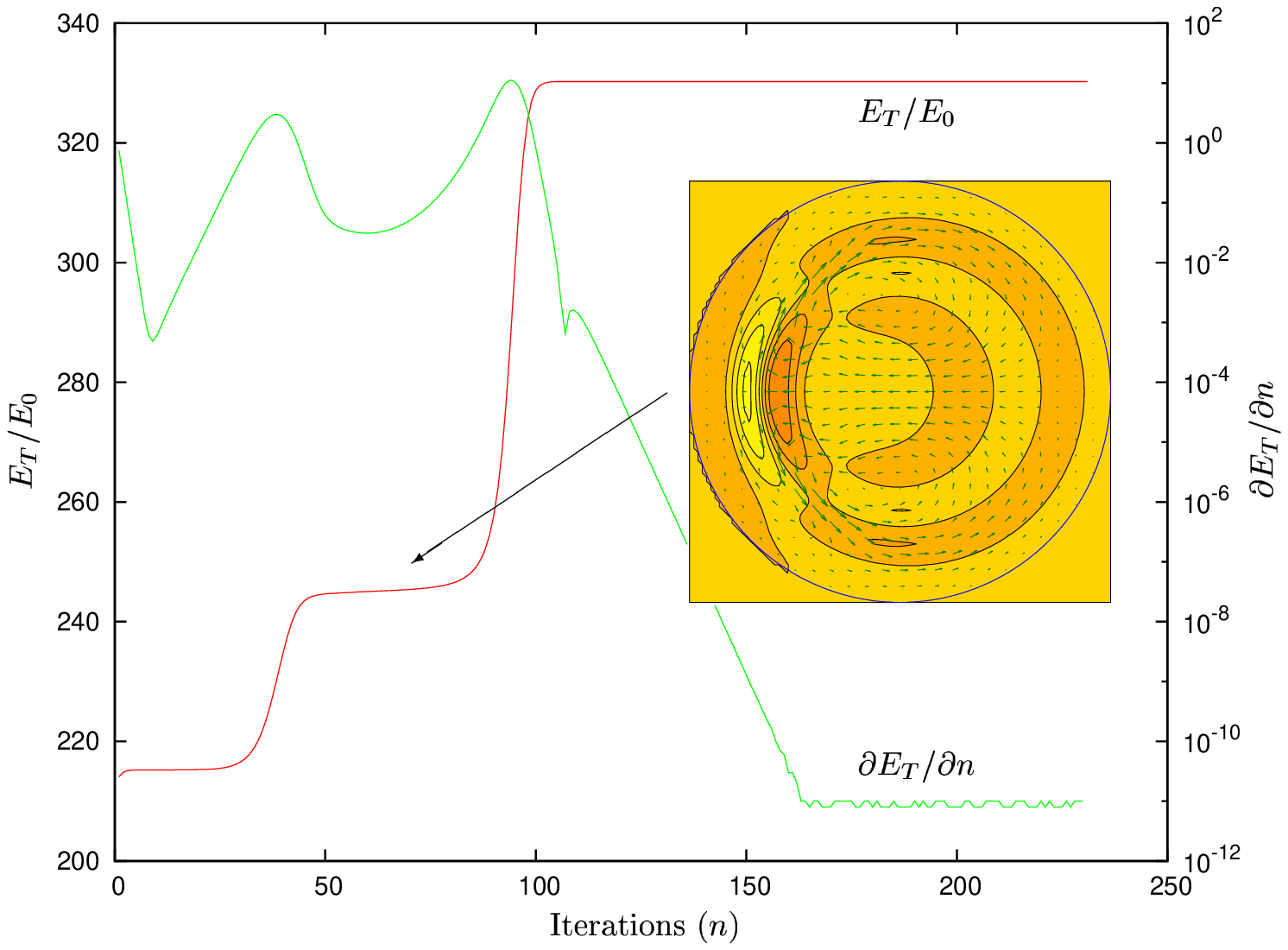}
  \caption[Convergence of the nonlinear optimal]{$Re=1750, E_0=2\times
    10^{-5}$. The iterations are seeded with a noisy version of the 2D
    optimal which converges to the 3D optimal by way of an intermediate
    `saddle' state (shown).\label{fig:saddle3_conv}}
% \end{center}
\end{figure}

\begin{figure}
% \begin{center}
\centering
  \includegraphics[width=\columnwidth]{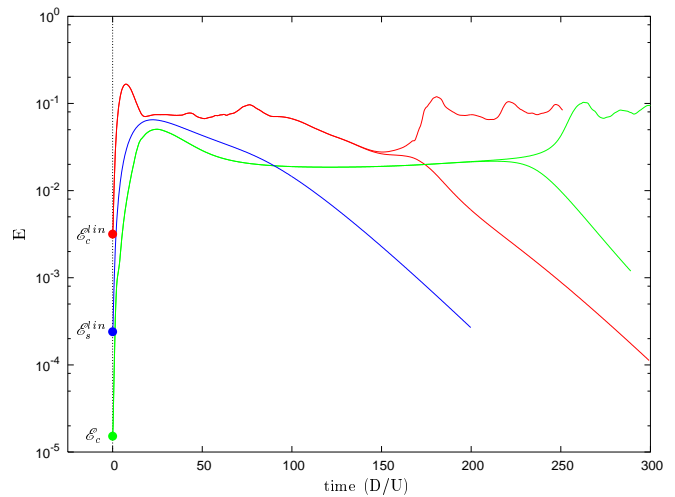}
  \caption{Re=2500. The green (lowest at $t=50\,D/U$) line shows the
    evolution of the 3D optimal when given initial energy
    $\mathscr{E}_c$. Because it is on the laminar-turbulent boundary
    two nearly identical initial conditions diverge after, in this
    case, $220D/U$. The blue (middle at $t=50\,D/U$) line is the
    evolution of the 2D optimal for the exact initial energy
    $\mathscr{E}_s^{lin}$ for which the streaks become linearly
    unstable.  The red (upper at $t=50 \,D/U$) line shows the 2D
    optimal given initial energy $\mathscr{E}_c^{lin}$ and allowed to
    evolve until it reaches a maximum amplitude whereupon $0.1\%$ by
    amplitude unstable perturbation is added. Again the
    laminar-turbulent boundary can be identified.\label{edges}}
% \end{center}
\end{figure}

For $E_0>2 \times 10^{-5}$ at $1750$ and $6.25 \times 10^{-6}$ at
$2250$, the iterative procedure fails to converge. In either case, 
a direct numerical simulation
starting with the 3D optimal at the highest energy value yielding
convergence does not reveal a turbulent episode. This implies that the
critical energy level, $\mathscr{E}_c(Re)$, for transition has not been
reached \footnote{There is the tacit assumption that the optimisation
  algorithm samples all possible flows of a given energy and thus if
  it converges smoothly, turbulence cannot be triggered at this energy
  level.}.  The reasons for this energy `gap' are unclear and leaves
open the possibility that a further new optimal may emerge. It is worth 
noting that the end state of the 3D nonlinear optimal decays more quickly 
that of the 2D optimal (which is in fact a close approximation of 
the least decaying eigenmode). Increasing the optimisation time, $T$, 
will therefore lead to the recovery of the 2D optimal at larger values 
of $E_0$. We then expect to able to converge onto the 3D optimal at 
correspondingly larger values of $E_0$, allowing the gap from $E_{3D}$ 
to $\mathscr{E}_c$ to be closed. 

What can be tested, however, is whether the 3D optimal is more efficient at
triggering turbulence than the linear optimal when rescaled. Taking
the initial condition
$A\mathbf{u}_{3D}(\mathbf{x};2\times10^{-5},1750)$, we gradually
increase the rescaling factor $A$ until $E_0=\mathscr{E}_c(Re)$ is reached.
Calculating the corresponding quantity for the linear optimal turns
out to be less clearly defined because some 3D noise is needed to
trigger turbulence. As a result we make 2 different estimates, one
strictly conservative and the other more realistic. The first
$\mathscr{E}_s^{lin}$ is obtained by taking
$A\mathbf{u}_{lin}(\mathbf{x};1750)$ and finding the initial energy
for which the resultant streaks are just linearly unstable in this
periodic domain \cite{zikanov,reddy}.  In the second $\mathscr{E}_c^{lin}$,
the same initial condition was used but $0.1\%$ of the most unstable
perturbation (as found from the previous computation) is added to the
streaks when they reach maximum amplitude. $\mathscr{E}_s^{lin}$
should be a (low) conservative estimate but even this is O(10) times
larger than $\mathscr{E}_c$ at $Re=2500$ - see figure \ref{edges} -
whereas the more realistic $\mathscr{E}_c^{lin}$ is O(100) times
larger.

\begin{figure}
% \begin{center}
\centering
  \includegraphics[width=\columnwidth]{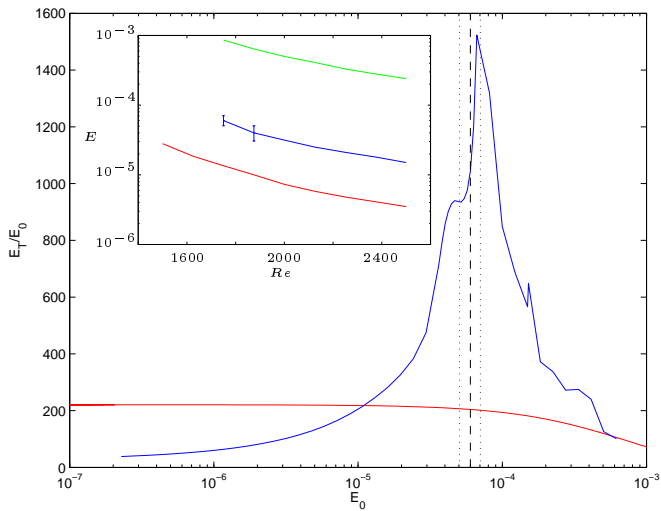}
  \caption{The effect of the initial energy on the growth of
    $A\mathbf{u}_{lin}$ (red) and $A\mathbf{u}_{3D}(\mathbf{x},2\times
    10^{-5},1750)$ (blue) at $Re=1750$. For small $E_0$ the 2D result
    is the optimal but after $E_{3D}=1.35 \times 10^{-5}$, the 3D
    optimal takes over. The vertical dashed line corresponds to
    $\mathscr{E}_c$, with the dotted lines being the relevant
    errorbars.  \textbf{Inset:} The dependence of $E_{3D}$ (red),
    $\mathscr{E}_{c}$ (blue) and $\mathscr{E}_s^{lin}$ (green) on $Re$
    ($\mathscr{E}_c^{lin}$ is even higher). For $Re < 2000$, errorbars
    on $\mathscr{E}_{c}$ indicate the energy range over which short to 
    to \emph{extended} turbulent episodes are triggered.\label{amplitude}}
% \end{center}
\end{figure}

In figure \ref{amplitude}(inset) we plot $E_{3D}$, $\mathscr{E}_c$ and
$\mathscr{E}_s^{lin}$ as a function of $Re$ which emphasizes that the
2D optimal (for which the linear result is an excellent approximation)
ceases to be a global maximum at an energy (at least) several orders
of magnitude before it approaches the laminar-turbulent boundary.  The
3D optimal, in contrast, crosses the laminar-turbulent boundary only
shortly after it emerges at $E_{3D}$ (e.g. at $\approx 5 \times
10^{-5}$ where $E_{3D}=1.35 \times 10^{-5}$ at
$Re=1750$). This means that the energy growth experienced by the 3D
optimal must increase dramatically with $E_0$ which is illustrated in
figure \ref{amplitude} at $Re=1750$: this growth is now a lower
bound on the maximum possible for $E_0>2 \times 10^{-5}$.
Assuming that a 3D optimal will always appear at subcritical energies 
(reasonable as the 2D disturbance cannot trigger turbulence), 
the critical energy must be  bounded from below by
$E_{3D}$ and above by $\mathscr{E}_{c}$.

%
% Discussion
%

In this letter we present the first demonstration that including
nonlinearities in the problem of transient growth substantially
changes the form of the optimal at energies below that needed to
trigger turbulence. The significance of this result comes from the
fact that transient growth analysis is currently the only constructive
approach (albeit with assumptions) for identifying critical
disturbances beyond exhaustive searches over initial conditions. As a
result, the new 3D optimal found here supersedes the linear optimal as
our current best theoretical prediction for the most dangerous
disturbance in pipe flow.

There are two key directions for improving the result presented here:
performing a further growth maximisation over $T$ and adopting
larger, more realistic flow domains.  Both represent formidable
extensions even with today's computing power. After all, the
discovery of the first true nonlinear optimal has had to wait almost
two decades after the linear result was established in pipe flow.
In larger domains, we expect further localisation of the nonlinear
optimal since energy is defined as a {\em global} quantity whereas
nonlinearity is important wherever the velocity field is
\emph{locally} large. This strongly suggests that in a long pipe the
optimal should localise fully (i.e. in the axial direction as well)
which would make it an interesting focus for experiments.

\begin{acknowledgments}
  We thank the referees for their comments. The calculations
  in this paper were carried out at the Advanced Computing Research
  Centre, University of Bristol.
\end{acknowledgments}

\end{document}